\documentclass{aa}
\usepackage{graphics}
\usepackage{epsf}
\usepackage{psfig}

\def\etal{{\it et al.} }
\def\eg{{\it eg.} }
\def\ie{{\it ie.} }
\def\cf{{\it cf} }

\begin{document}

\thesaurus{03 (03.13.4);	
	   11 (11.03.1);	
	   12 (12.12.1)		
          }

\title{EISily looking for distant clusters of galaxies - 
a new algorithm and its application to the EIS-wide data}

\author {
C. Lobo
          \inst{1,2}
 \and 
A. Iovino
          \inst{1}
\and
D. Lazzati
          \inst{1}
\and
G. Chincarini
          \inst{1}
	}

\offprints{C. Lobo, lobo@oal.ul.pt}

\institute{
        Osservatorio Astronomico di Brera, via Brera 28,
	20121 Milano, Italy
         \and
	Observat\'orio Astron\'omico de Lisboa, Tapada da Ajuda,
	1349-018 Lisboa, Portugal
     }

\date{Received April 27, 2000; accepted month ??, 2000}

\maketitle
\markboth{Lobo \etal, EISily looking for distant clusters of galaxies}{}

\begin{abstract}
We present a new algorithm to search for distant clusters of galaxies
on catalogues deriving from imaging data, as those of the ESO Imaging
Survey. 

Our algorithm is a matched filter one, similar to
that adopted by Postman \etal 1996, aiming at identifying cluster
candidates by using positional and photometric data simultaneously.
The main novelty of our approach is that spatial and luminosity filter
are run separately on the catalogue and no assumption is made on the
typical size nor on the typical $M^{*}$ for clusters, as these
parameters intervene in our algorithm as typical angular scale
$\sigma$ and typical apparent magnitude $m^{*}$. Moreover we estimate
the background locally for each candidate, allowing us to overcome the
hazards of inhomogeneous datasets.  As a consequence our algorithm has
a lower contamination rate - without loss of completeness - in
comparison to other techniques, as tested through extensive
simulations. We provide catalogues of galaxy cluster candidates as the
result of applying our algorithm to the I--band data of the EIS-wide
patches A and B.


\end{abstract}

\section{Introduction}\label{introducao}


The quest for high--redshift ($z \ga 0.5$) clusters of galaxies has
recently received a lot of attention, and several search methods have
been put forward.  The mere existence of rich clusters of galaxies at
very high redshifts is cosmologically relevant and their importance as
a discriminant among different theoretical cosmological models is
widely acknowledged (see \eg Viana $\&$ Liddle \cite{viana}; Carlberg
\etal \cite{carl}; Bahcall, Fan $\&$ Cen \cite{bfc}; Bartelmann \etal
\cite{bartel}).  On the other hand, a detailed study of the properties
of clusters, as well as of their member galaxies, at high redshifts
and the confrontation of these with local well-known systems gives
precious insight on their formation and on the evolution of their
properties with redshift and environment (Dressler \etal \cite{dress};
van Dokkum \etal \cite{vandok98a}, \cite{vandok98b}; Smail
\etal \cite{smail98}; Morris \etal \cite{morris}; Stanford \etal \cite{stan98} 
and references therein). The now well-established evolution of galaxies with redshift
still lacks a clear understanding of the physical processes that guide
it: clusters are privileged observational targets to distinguish
between intrinsic and environmental effects.

To perform the search for distant clusters of galaxies, various
authors have been taking advantage of several windows in the spectral
emission of the different cluster components, namely galaxies and gas,
and applying the best suited method in each case. Thus, optical, near
IR, X--ray and even variations in the cosmic microwave background emission
have been used to detect cluster candidates at distant redshifts.  All
different methods have advantages and drawbacks and the key is to view
them as complementary. Describing each one of them and presenting
their respective pros and cons is beyond the scope of this
paper. Nevertheless, it is useful to recall what has been done up to
now, in order to place our work in the actual scenario of these
researches.

Among the different techniques that basically search for surface
density enhancements on the galaxy 2D spatial distribution - adding or
not magnitude information as well -, some of the most popular ones are
the application of adapted filters (\eg kernel, wavelets, matched
filters), counts-in-cells techniques, percolation algorithms or even,
lately, Voronoi tessellation, on optical photometric data (\eg
Schectman \cite{schectman}; Lumsden \etal \cite{lum}; Dalton \etal \cite{dalton}; 
Escalera $\&$ MacGillivray \cite{escalera95}, \cite{escalera96}; Pisani \cite{pisa}; 
Lidman $\&$ Peterson \cite{lid}; Fadda \etal \cite{fadda}; Postman \etal \cite{postman}; 
Gal \etal \cite{gal}; Ostrander \etal \cite{ostra}),
and on NIR data (Stanford \etal \cite{stan97}; Mendes de Oliveira \etal \cite{mendes},
Ramella, Nonino $\&$ Boschin \cite{ramella}). Apart from these more or less
elaborate techniques, simpler high-contrast methods still prove to be
successful (Couch \etal \cite{couch}), but they probably don't sample
adequately the full distant cluster population and their selection
criteria are not well defined.  Some of these algorithms benefit also
from - or are mainly based upon - multiband colour information that
helps isolating red galaxies at higher redshifts (Gladders \& Yee
\cite{glad}).

Very significant results have been achieved {\it via} the search for
sources of extended emission in X--ray surveys with detection
algorithms which are designed to probe a broad range of cluster
parameters such as X--ray flux, surface brightness and morphology (\eg Henry \etal
\cite{henry92} with EMSS; RIXOS by Castander \etal \cite{cast}; the RDCS by Rosati
\etal \cite{rosaetal95}, \cite{rosaetal98}; NEP by Henry \etal \cite{henry97}; 
Gioia \cite{gioia}; WARPS by Scharf
\etal \cite{scharf}; Jones, Scharf, Ebeling \etal \cite{jse}; SHARC by Collins \etal
\cite{collins}; the BCS by Ebeling \etal \cite{ebel}; the CfA large area survey by
Vikhlinin \etal \cite{vik98a}, \cite{vik98b}; REFLEX by B\"{o}hringer \etal \cite{bohr}).

Other emerging strategies for cluster search include: the detection of
extragalactic background light fluctuations in shallow optical images
(Dalcanton \cite{dalcanton}; Zaritsky \etal \cite{zar}); deep imaging around privileged
sites of density enhancements such as distant powerful radio-galaxies
or radio-loud quasars (\eg Le F\`evre \etal \cite{lef}; Deltorn \etal \cite{deltorn});
narrow--band imaging to search for concentrations of Ly--$\alpha$
emitters around previously known weak radio QSO's at high redshift
(\eg Pascarelle \etal \cite{pasca}; Campos \etal \cite{campos}).  Recently, the
detection of decrements in the cosmic microwave background radio
emission were also attributed to the presence of distant gaseous
systems, possibly clusters (Jones, Saunders, Baker \etal \cite{jsb};
Richards \etal \cite{rich}), that scatter the microwave background radiation
{\it via} the Sunyaev--Zel'dovich effect (Sunyaev $\&$ Zel'dovich
\cite{suny}). One should mention, though, that these strategies for cluster
search refer to $z > 2$ systems and we ignore {\it ``whether these are
massive, collapsed systems, groupings within unvirialized ``sheets''
of galaxies, or collections of ``protogalactic'' fragments destined to
merge into single, more massive galaxies''} (Dickinson \cite{dick}).\\

If one aims at producing statistically significant results, a good
catalogue of clusters, preferably spanning a large interval of
redshift, obtained using well defined selection criteria, over a
reasonable sky area, is needed.  Our algorithm, by
uniformly detecting clusters over a wide range of redshifts and
cluster sizes, is well suited to provide such a catalogue. We stress
that it does not aim, however, at performing accurate estimates of
neither the redshift nor the richness of the cluster candidates.

The ESO Imaging Survey (EIS, Renzini $\&$ da Costa \cite{renzini}), covering a
final area of $17$ square degrees of the southern sky in the I-band,
up to limiting magnitude $I \sim 23 $, provided us with a good
opportunity to produce such a catalogue. We have developed an
automated cluster search algorithm and applied it to the catalogues
derived from the EIS imaging data to obtain a reliable set of
cluster candidates up to estimated $z \sim 1.1$.  Having a list of
robust candidates is highly desirable, before proceeding to the
spectroscopic observations with very large telescopes, and our
algorithm has a high success rate - that is, a high completeness level
- without being overwhelmed by contamination in the form of spurious
detections. We have already selected three of the highest redshift
candidates ($z \ge 0.5$) and performed the spectroscopic follow-up
with VLT, confirming their real existence as physically bound
systems (discarding the possibility of false chance alignments) and
determined their distance/redshift with accuracy.  

In this paper we describe our algorithm in section \ref{algoritmo},
stressing its strong points and advantages relatively to others
present in the literature. Applying it to the EIS-wide data of patches
A and B produced a catalogue of candidate clusters of galaxies that we
present in section \ref{catalogos}. This set is quite different from
the one produced by the EIS--team of ESO for the same original data set but
using a different approach and in section \ref{compolsen} we fully
investigate all possibilities that may account for such a
discrepancy. Final remarks are given in section \ref{fim}.

We shall use $H_0 = 50$ km~s$^{-1}$Mpc$^{-1}$ and $q_0 = 0.5$
throughout this paper, unless explicitly stated otherwise.

\section{Our new algorithm}\label{algoritmo}


In their pioneering work on the Palomar Distant Cluster Survey,
Postman and collaborators (Postman \etal \cite{postman}, P96 hereafter) wrote a
matched filter algorithm to identify cluster candidates by using
positional and photometric data simultaneously. Slight variants have
been proposed hence, with differences lying in some details of the
detection processes (Kawasaki \etal \cite{kawa}) or in the generalization of
the algorithm so as to render it applicable to any type of survey data
(Kepner \etal \cite{kep}).  Our work, instead, tries to improve on the P96
algorithm in the sense of removing the {\it a priori} assumptions
that are implicit in the Postman \etal technique.

The P96 algorithm relies on the choice of both a given cluster profile
- modified--Hubble or King--like (hereafter often referred to as King
filter for simplicity) - with a typical cluster scale - the core
radius $r_c$ - and a cut--off radius, and of a typical $M^*$ - the
chosen parameterization of the Schechter function. Both quantities,
$r_c$ and $M^*$, are {\bf rigidly coupled} to detect cluster
candidates and assign them a tentative redshift.
Each one of these quantities being a function of
distance (and thus cosmology), their implementation on any algorithm
implies a dependence on the adopted cosmology and on the chosen amount
of evolution, not to mention the dependence on the particular values
actually chosen for $r_c$ and $M^*$.

In our new algorithm {\bf the spatial and luminosity part of the
filter are run separately and successively} on the catalogue, with
{\bf no assumption} on the typical size or typical $M^*$ for clusters.
In fact, these parameters intervene in our algorithm only as a typical
angular scale and typical apparent magnitude $m^*$, bearing no ties to
fixed physical scales nor to absolute magnitudes through redshift
dependence.  This has the consequence of removing the need for a
choice of fixed physical values for these two quantities, and for a
choice of evolutionary models.  Moreover, the fact of
not coupling the space and the luminosity parameters also enables us
to reach higher values of completeness: a candidate can be retrieved
even if it does not flag a maximum likelihood at the very same
redshift value simultaneously for both the space and the luminosity
distributions (a situation that would lower its global likelihood when
using the P96 algorithm).

One further advantage consists in the {\bf local estimate of the
background} for each cluster candidate. In the approach of P96, the
magnitude distribution is assumed constant all over the catalogue and
so is the background spatial density.  This may be hazardous
especially in what concerns the spatial part, where local
inhomogeneities may hamper the detection in shallower regions of the
catalogue. Due to varying observing conditions, a part of the EIS
data, at least, is reported to be non uniform (Olsen \etal \cite{olsenb}), so
this feature of our algorithm is of great help, as it allows us to
tackle the problems of inhomogeneous data sets and to achieve higher
levels of completeness.\\

In our algorithm local enhancements in the projected galaxy density
are first selected through the Gaussian filtering, and a ``spatial
probability'' of them being spurious is computed. Subsequently, the
maximum-likelihood ``filtering'' on the apparent magnitudes searches
for the presence of a Schechter distribution superimposed on the local
background (which follows a power--law instead). This second step
leads to the assessment of a corresponding ``luminosity probability''
that, multiplied by the spatial one, produces the final probability of
each candidate being a spurious one.  This means that the lower this
value, the more confident the candidate.  Notice also that this final
quantity is always lower than either one of the partial spatial or
luminosity probabilities as it is their product. \\

We will now discuss in detail the components of the algorithm.

\subsection{The spatial filter}\label{spatial}

In the spatial part we have chosen to work with a {\bf Gaussian
filter}.  This choice, while avoiding to specify too particular a
shape for the profile of the cluster candidates, brings along all the
advantages and favorable mathematical properties of the Gauss
function. It is worth remembering that the gaussian function in
Fourier space is more compact than the King function. As a consequence
the convolution with a random distribution of galaxies (a white noise
in fourier space) will produce less spurious detections for the Gauss
filter than for the King filter. 

Regarding completeness, it can be interesting to show, using
simulations, the relative advantages of the Gaussian filter with
respect to the King filter.  A simple way of doing it is to simulate
an area with randomly distributed field galaxies where to embed
clusters, and apply to this area the two spatial filters in question
to perform cluster detection and to compare directly completeness for
both of them.

Comparing the two filters is not straightforward, though: their shapes
are intrinsically different so that there is no direct correspondence
between the respective typical scales ($r_c$ and $\sigma_{ang}$).  The
King--like profile has a slimmer central peak with broad wings while the
Gaussian shows the well known bell--shaped form.  However, when
cross-correlating the two filters, the highest signal is achieved if
we consider $\sigma_{ang} \sim 2.52 \times r_c$, suggesting that a
factor of $ 2.52$ should be used in the simulations when comparing the
relative efficiency of the Gauss {\it versus} King filters. Besides,
since when searching for a spatial density enhancement one has to move in
steps of dimension comparable to the typical scale of the filter, this
also implies that the Gaussian filtering requires a factor of
$(2.52)^2 \simeq 6$ less search points with respect to the King
filtering.

As we have already mentioned, in our search we will not fix any
typical physical size for our filter, but we will use, instead, a
range of angular sizes $\sigma_{ang}$, namely from $\sim 0.35$ up to
$\sim 1.42$ arcmin in five steps of ratio $\sqrt2$.  These values were
chosen bearing in mind the range of reasonable dimensions of the
cluster candidates spanning the redshift interval that we expect to
probe with the EIS data ($z \la 1.2$). The corresponding five $r_c$
values to be used in the spatial search with the King--like filter are
those obtained using the relationship $\sigma_{ang} \sim 2.52 r_c$.

\subsubsection{Simulations}\label{simspatial}

The simulations performed for this comparison were
done as a function of the signal-to-noise threshold chosen, of the
cluster profile, richness and redshift.

As the final goal was to apply our algorithm to the EIS data, to
simulate the galaxy background distribution we used the data from the
EIS itself (see section \ref{catalogos}), by random shuffling galaxy
positions and magnitudes within the limits of the survey.

The contamination rates were assessed by running the Gaussian filter and
the King filter on these pure background fields.

In what regards completeness, this rate was assessed by running the
two filters on the background fields plus cluster galaxies.  This
means that, for each simulation, a cluster of galaxies with different
characteristics was placed randomly within each background frame, with
the only constraint of being at adequate distance from the frame
borders (\ie a distance larger than both 10 times the $\sigma_{ang}$
used for the search as well as 5 times the angular size of the
embedded cluster), in order to avoid border effects. The
characteristics of the clusters were defined according to the
following prescription:

\begin{itemize}
\item [1.] The cluster mean redshift: ranging from $z = 0.2$ till 
$z = 1.2$ at intervals of $\Delta z = 0.2$.

\item [2.] The surface density profile: a power law in radius of the
form $r^{\beta}$. The index $\beta$ was set to three different values
($0$, $-1$ and $-2$) when testing the spatial filter only (as is the
case of this section). But it was fixed to the typical in--between case
$\beta = -1.4$ for the overall simulations (further ahead, in section
\ref{overallcomplete}, when we will apply the complete algorithm - that
 is, both spatial and luminosity filters - to the mock catalogues); this 
provided a compromise that also allowed a direct
comparison with the P96 results. Furthermore, we introduced a small
central smoothing region in the profile, with the typical size of a
cluster cD galaxy ($r_{smooth} \sim 35$ kpc, the average diameter of
the two central giant galaxies in Coma, as provided by the
NED-database), to avoid a cusp. Both $r_{smooth}$ and the total radial
size of the cluster, $1$ Mpc, were translated into the corresponding
angular sizes according to the redshift chosen in item 1. Doubling the
cluster size to $2$ Mpc did not change the results of our simulations
in terms of completeness.

\item [3.] The luminosity function - a Schechter, which is generally
adopted to describe the luminosity distribution of cluster galaxies
(Schechter \cite{schechter}), but with parameters determined by
Colless (\cite{colless}) for a set of $14$ clusters of mean redshift
$<z> \simeq 0.0851$ observed in the $B_J$ band. Thus, the faint end
slope is $\alpha = -1.25$ and the characteristic apparent magnitude,
at that mean redshift, is $m_I^* = 14.8$ when converted to the I-band
(Fukugita \etal \cite{fuku}). The $m_I^*$ used in each simulation was
changed according to the redshift (item 1) and to the k-correction
applied (see item 5 below).

\item [4.]  The richness, as given by $n^*$, the number of galaxies
brighter than the characteristic magnitude $m^*$. Three cases were
considered: $n^* = 30, 50$ and $80$ (following Schechter
\cite{schechter}). These can be roughly identified with Abell richness 
classes $R = 1$, $2$ and $3$ (Abell \cite{abell}; see also Bahcall 
\cite{bahcall88}).

\item [5.]  The dominant morphological type, as implied by the
k-correction that is used. Actually only one extreme case was tested:
clusters entirely composed of elliptical galaxies.  We deliberately
chose to ignore both the evolutionary corrections and the possibility
of having later morphological types as the cluster dominant
population, since this would correspond to a more favorable scenario
of rendering cluster members brighter, which would only facilitate
their detection. k-corrections have been provided by the
spectro-photometric evolutionary model {\it PEGASE} of Fioc $\&$
Rocca-Volmerange (\cite{fioc}) tuned to the cosmology parameters adopted.

\end{itemize}

Once these characteristics are fixed, we can compute, for each
cluster, the number of member galaxies that are observable within the
apparent magnitude limit. Positions and magnitudes are next assigned
to these cluster galaxies by random shots drawn from items 2 and 3,
respectively.

This was done in sets of $100$ simulations and the two spatial filters
were applied to these mock catalogues. Both field and cluster galaxies
were limited to a magnitude cut--off of I=$22.0$, the one we adopted
also for the EIS data.\\

\begin{figure}[htbp]
\epsfysize=9cm
\epsfbox{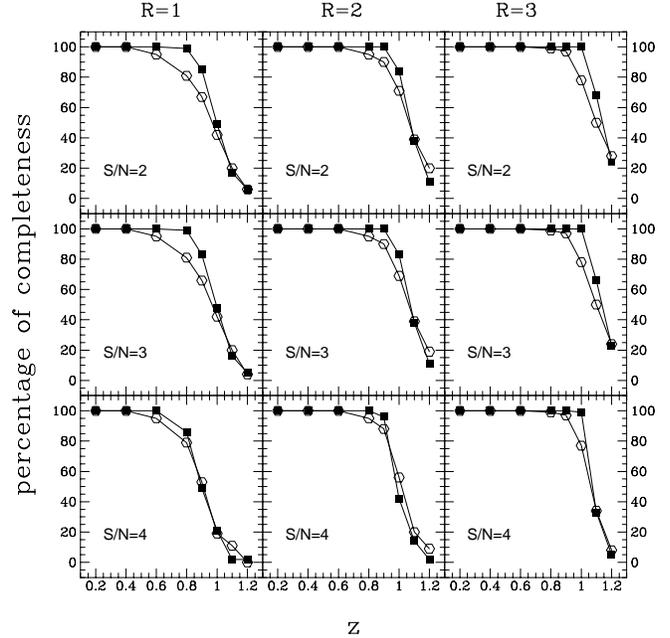}
\caption{Comparative performance, in completeness rate, between the
spatial filter used in our new algorithm with the five adopted
angular widths ($\sigma_{ang}$) - filled squares - and the P96 King--like
spatial filter with the corresponding five core radii - empty circles
- such that $r_c = 1/2.5 \sigma_{ang}$.  Each point represents the
mean value of a set of $100$ bootstrap simulations. Clusters are
elliptical dominated, have radial profile $r^{-1}$ and richness class
$R \sim 1$ (left), $R \sim 2$ (center), and $R \sim 3$ (right). See
text for further details.}
\label{fig:GKprofilepm1}
\end{figure}

\begin{figure}[htbp]
\epsfysize=9cm
\epsfbox{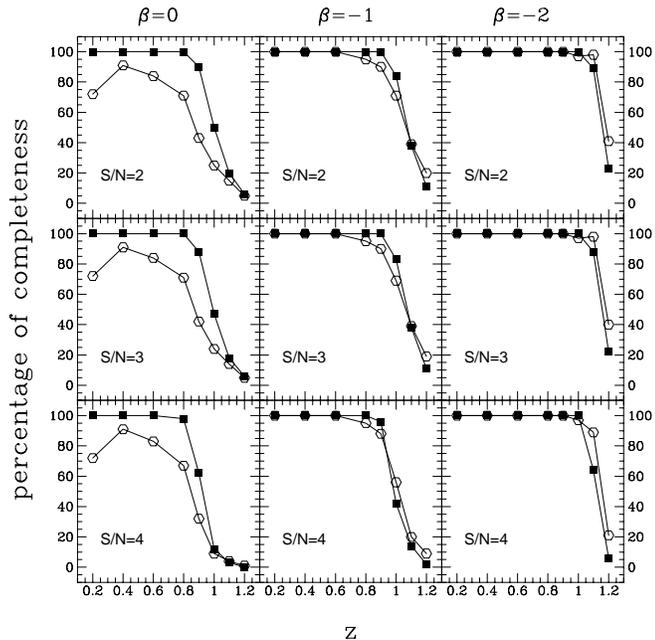}
\caption{Same as figure \ref{fig:GKprofilepm1} but now the richness class 
has been fixed to $R = 2$ and different $r^{\beta}$ radial 
profiles are compared in terms of filter completeness performance. Left 
panels represent $\beta = 0$ clusters, in the middle stand $\beta = -1$ 
clusters and the right panels show results for $\beta = -2$, steeper profile, 
clusters.}
\label{fig:GKrichnessR2}
\end{figure}

In terms of completeness, the results are summarized in figures
\ref{fig:GKprofilepm1} and \ref{fig:GKrichnessR2}.

According to the plots in figures \ref{fig:GKprofilepm1} and
\ref{fig:GKrichnessR2}, the comparative performance in completeness
rate shows that the Gaussian filter is more efficient in most of the
cases than the King filter, and this is especially true in particular
in the difficult case when the cluster profile is less peaked.

\subsection{The luminosity filter}

To assess the signal given by the luminosity information we apply a
maximum-likelihood technique on the apparent magnitude distribution of
the candidates previously found with the spatial filter.  We use as
functional form for the luminosity function a Schechter (\cite{schechter})
function expressed in apparent magnitudes, thus avoiding the P96
choice of an intrinsic $M^*$ and the choice of a given cosmology
($H_0$ and $q_0$) and morphological content (translated in the
assumption of a given k-correction and, eventually, of an e-correction
as well). Also, unlike P96, we have decided to implement the
generalized treatment for this likelihood filter, following Schuecker
$\&$ B\"{o}hringer (\cite{schuecker}). This complete treatment uses the exact
mathematical equations, allowing to keep track of all possible errors
that would affect any eventual direct redshift estimate (like the one
performed by P96).  It does not add major computational effort nor
time. Besides, it also renders unnecessary the final {\it ``cluster
signal correction''} demanded by the complex approximations of the P96
procedure.

We thus took equation (8) of Schuecker $\&$ B\"{o}hringer
(\cite{schuecker}) and adapted it to compute the likelihood for this
luminosity part of our algorithm. Notice that, as we decouple it from
the spatial part, the galaxy distribution profile is suppressed from
the original formula so that the final log-likelihood ratio, for a
given $m^*$, is:

\begin{eqnarray}\label{eq:like}
\ln\left(\frac{L_{S+bkg}}{L_{bkg}}\right)\,= \sum_{i=1}^N\,\ln
\left[\,\frac{N\,-\,\Lambda}{N}\,\left(1+
\frac{\Lambda\,\phi(m_i-m^*)}
{\left(\,N\,-\,\Lambda\right)\,b(m_i)}\,\right)\,\right]\nonumber
\end{eqnarray}
\hspace{8.cm}(1)
\setcounter{equation}{1}
\vspace{0.5cm}

\noindent where $L$ stands for the likelihood function, $S$ and
$\phi(m_i-m^*)$ refer to the Schechter function parameterized by a
given $m^*$, $bkg$ stands for the background galaxies and $b(m)$ for
the differential magnitude number counts of the background galaxies;
$b(m)$ is mathematically described by the fit to the magnitude
distribution of all galaxies in the catalogue. N is the total number
of cluster and field galaxies present inside a $2.5 \sigma_{ang}$
radius circle, where we also estimate a coarse richness parameter
$\Lambda$ (as will be detailed in section \ref{descricao_algoritmo}).

Note that $\phi$ has been normalized, \ie:

\begin{equation}\label{intS}
\int_0^{m_{lim}}\,\phi(m-m^*)\,dm\,=\,1
\end{equation}

\noindent $m_{lim}$ being the limiting magnitude of the catalogue.

For each candidate we compute the log-likelihood ratio (equation
\ref{eq:like}) for different $m^*$ values, set within the limits of
magnitude of the galaxy catalogue by steps of $0.1$ magnitude. The
slope of the Schechter function is fixed to $\alpha = -1.25$,
according to the mean value derived by Colless (\cite{colless}) and in agreement
with typical cluster luminosity function parameters found in the
literature since the work of Schechter (\cite{schechter}). By identifying the
absolute maximum in the distribution of the log-likelihood ratios
relative to all $m^*$, we automatically select the output $m^*$ for
that candidate.

\subsection{Detailed description of the algorithm steps}\label{descricao_algoritmo}

In this section we present a detailed description of the way the
algorithm works when applied to the data. 

We note that we shall often interchange
probability with the complementary Gaussian percentile and the
associated level of standard deviations, referred to as $S/N$ throughout
this paper. This approximation was used also in P96 so, by adopting it
as well, we render comparisons among our results easier.

The algorithm starts off by reading the coordinates and magnitudes of
the galaxies present in a magnitude limited catalogue. Then, for each
one of the five selected angular apertures $\sigma_{ang}$ of the
Gaussian, the code:

\begin{itemize}
\item[$1.$] Builds the search grid for the spatial filter analysis. The 
spacing between adjacent grid points is $\sigma_{ang}$ but no grid point is 
allowed within $3 \sigma_{ang}$ of the catalogue's borders in right ascension nor in declination.

\item[$2.$] Computes a first hand estimate of the local background
density associated to each grid point by counting the number of
galaxies in a circle with a $6 \sigma_{ang}$ radius centered at the
grid point in question. (All cases of grid points distant less than $6
\sigma_{ang}$ from the catalogue's borders are obviously corrected by
shifting the circle.) Notice that, in this way, the background spatial
density is not assumed constant all over the catalogue, but is
estimated around each point. This characteristic proves to be
particularly efficient in dealing with data - like those of EIS in at
least some of the patches - where varying observing conditions result
in non homogeneity of the galaxy catalogue (Olsen \etal \cite{olsenb}).

\item[$3.$] Performs the Gaussian filtering at each grid point
considering all galaxies within $3 \sigma_{ang}$ from it and weighing
their relative position through the Gauss function.  The resulting
output signal is normalized (respective mean subtracted and
division by the corresponding standard deviation), obtaining the $S/N$ 
appropriate for each grid point. 

\item[$4.$] Selects all grid points which flag a local maximum, \ie,
for which the spatial signal is greater than that of all their
immediate grid neighbours. No points at the extremes of the grid are
allowed in this choice, thus avoiding "border effects". The signal
must also be above a given threshold, imposed as an input in the
programme.  We set this spatial detection threshold to $2.5$ (the
equivalent to the $98$th percentile).

\item[$5.$] For these maxima, our algorithm proceeds to the fine
centrage: the Gaussian filtering is redone, maintaining the filter
angular size, but building a finer grid in an iterative procedure
(down till an eighth of $\sigma_{ang}$) around the tentative
candidate. The fine tuned center is finally chosen as the position
where the spatial signal is maximized in the ensemble of the points of
this denser grid.

\item[$6.$] The five different scale catalogues are cross-correlated
in order to remove double detections. Detections are considered double
if the distance between their centers is lower than or equal to the
mean of their scales. In this case, only the highest signal detection
is kept.

\item[$7.$] For each selected candidate, the next step is to obtain a
preliminary estimate of its richness, $\Lambda$, to be used in the
computations of the likelihood of the luminosity part (see equation
\ref{eq:like}).  $\Lambda$ is the approximate number of galaxies
statistically belonging to the cluster candidate.  We obtain it by
subtracting to the N galaxies inside a $2.5 \sigma_{ang}$ radius the
local background, estimated in a annulus around each candidate.  This
ring has an inner radius of $5 \sigma_{ang}$ and its area totals
$0.06$ deg$^2$. These values were chosen after evaluating the galaxy
density distribution around several candidates in the final EIS
catalogue.

\item[$8.$] For each candidate we compute the log-likelihood ratio
(equation \ref{eq:like}) for the set of $m^*$ values and determine the
one that maximizes it. The probability associated to this value is
obtained by bootstrap: the catalogue is randomized, the spatial filter
is run on it and the distribution of the log-likelihood for the
flagged spatial detections is computed thus allowing an estimate of
the probability of having an $L_{max}$ greater or equal than that
measured for our candidate.

\item[$9.$]At this point we possess the $\sigma_{ang}$ and $m^*$ that
independently maximize both the spatial and the luminosity signals and
their respective significance, $P_{space}$ and $P_{lum}$. We can
combine these probabilities to obtain a final global
probability, $P_{tot}$, for the detection of each cluster candidate.
This is simply done by multiplying $P_{space}$ by $P_{lum}$.
\end{itemize}

\begin{figure}[htbp]
\epsfysize=9cm
\epsfbox{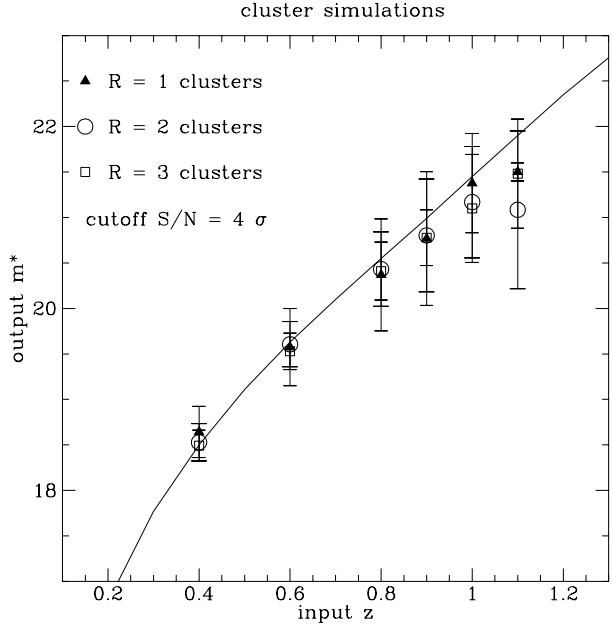}
\caption{Results from the simulations (that allowed us to assess the overall 
completeness rate - see section \ref{overallcomplete}) showing the comparison 
between the redshift value of the simulated clusters and the output $m^*$ 
obtained from our algorithm. The curve shows the relationship between $z$ and 
$m^*$ used to define the input $m^*$ in our simulations (see text for further 
explanations). Each point is the mean of 100 simulations. $1 \sigma$ error 
bars are drawn. 
}
\label{fig:msz}
\end{figure}

\subsection{Richness estimate}\label{rich}

The estimate of the richness of our cluster candidates is an important
step, as we would like to quantify the percentage of groups {\it
versus} rich clusters that we will be detecting.

Whichever richness estimate one decides to adopt, a necessary step is
the determination of a physical radius for each cluster candidate
which, in turn, requires the estimate of its redshift.  Unlike P96,
our method does not provide a direct estimate of the redshift for our
candidates. We obtained such estimate using $m^*$ and its
relationship to $M^*$, whose adopted value is that reported by Colless
(\cite{colless}), to be redshifted and corrected using the k-corrections typical
of ellipticals, as determined by Fioc $\&$ Rocca-Volmerange (\cite{fioc}).
Simulations show that the value of $m^*$ obtained through our
algorithm retrieves well the $M^*$ introduced in the simulations - see
figure \ref{fig:msz}.  On the contrary, the value of $\sigma_{ang}$
retrieved by our algorithm does not correlate well with redshift: the
angular size chosen by the algorithm has a large scatter with respect
to the physical size of the clusters entered in the simulations.

Having thus obtained a rough estimate of the candidate's redshift, we
can also estimate its richness, adopting the $N_{0.5}$ parameter
proposed by Bahcall (\cite{bahcall81}).  It consists in counting the number of
cluster member galaxies brighter than $m_3 + 2$ (where $m_3$ is the
magnitude of the third brightest cluster galaxy), located within a
projected radius of $0.5$ Mpc from the cluster center, the typical
size of a cluster X--ray emitting region.  Background correction is
estimated from similar counts performed in the entire catalogue
region, using $0.2$ magnitude bins. $N_{0.5}$ was found to be well
correlated with the system's velocity dispersion $v_r$ by Bahcall
(\cite{bahcall81}, see also \eg Alonso \etal \cite{alonso} for further studies on this type
of correlation) for a sample of $26$ nearby clusters ranging from rich
systems ($v_r \sim 1500 km s^{-1}$) to small groups of galaxies ($v_r
\sim 100 km s^{-1}$) and has already been used in the literature for
richness estimates of distant clusters (Deltorn \etal \cite{deltorn}).  It has
the advantage of avoiding the higher uncertainties in background
subtraction present when adopting the standard Abell (\cite{abell}) estimate,
which is performed within a radius of $3$ Mpc (see \eg P96 for
comments on this). In fact, $N_{0.5}$ is relatively insensitive to
uncertainties in either the background correction or in the exact
position of the cluster center (Bahcall \cite{bahcall81}). $N_{0.5}$ can be
translated into the typical Abell richness classes using the
indicative relation given by Bahcall (\cite{bahcall81}), $N_{Abell} \simeq 3.3
\times N_{0.5}$, and then checking table 1 of Bahcall (\cite{bahcall88}). This
means that \eg $N_{0.5} < 15$ corresponds to Abell richness class $R
\sim 0$, while $N_{0.5} > 24$ is equivalent to $R \ga 2$.

\subsection{Expected final overall rates of completeness and contamination -- simulations}\label{overallcomplete}

We ran a new set of extensive simulations in order
to assess the overall efficiency of the algorithm. The layout of these
simulations being the same as the ones that were used for estimating
the completeness and contamination rates of the spatial filter alone
(see section \ref{simspatial}), we shall not repeat their
details. Here, it will be enough to mention that the mock catalogues
(pure field for contamination rates or field plus cluster galaxies for
completeness purposes) were now submitted to the algorithm, and not
only to its spatial filter as had been done in section
\ref{simspatial}.\\ 

\begin{figure}[htbp]
\epsfysize=9cm
\epsfbox{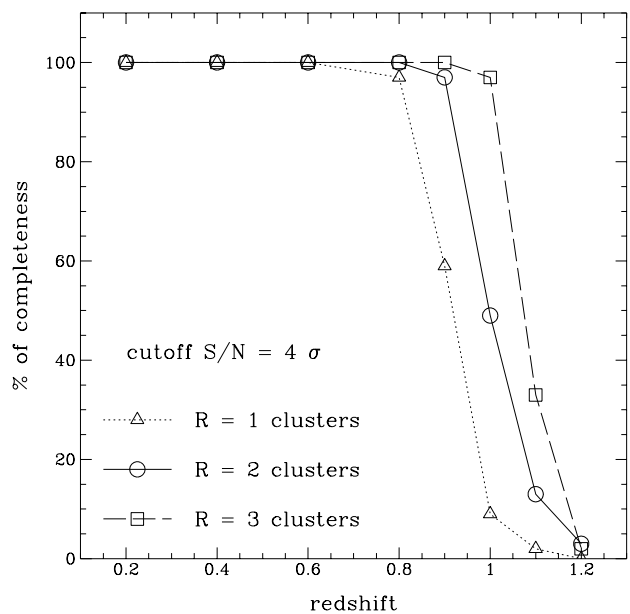}
\caption{Percentage of completeness for Abell richness classes $R = 1$ (poor), 
$R = 2$ (intermediate) and $R = 3$ (rich) clusters as established through 
simulations. Clusters have a power law radial profile of type r$^{-1.4}$, 
are dominated by ellipticals, and have a typical luminosity function as 
determined by Colless (1989). We applied elliptical k-corrections (Fioc $\&$ 
Rocca-Volmerange 1997) to cluster galaxies.} 
\label{fig:completeness}
\end{figure}

The results regarding the overall completeness rates are shown in
figure \ref{fig:completeness}. Even setting the detection threshold at
$S/N \sim 4.0$ - which renders contamination rates negligible (see
below for the final overall rate of contamination directly assessed on
the EIS data) - we are able to achieve a completeness of $\sim 95\%$
until $z \sim 0.9$ for richness 2 Coma-like clusters, and of $\sim
97\%$ up to $z \sim 0.8$ for poorer richness class 1 systems. For $R =
3$ clusters such high values of completeness hold even in the redshift
bin $z \sim 1.0$, always for the no-evolution case. These results
improve on the values obtained using the P96 algorithm (see
their figure $20$). In fact, while we set the Schechter slope of the
simulated clusters to a steeper value ($-1.25$ instead of $-1.1$), the
fact of having adopted a brighter magnitude limit ($22.0$ {\it versus}
P96's $22.5$)
renders a direct comparison of both works plausible. We checked this
by integrating both Schechters for the two magnitude limits, thus
getting 38(us) {\it versus} 50(P96) galaxies for R=1 clusters, 63 {\it
versus} 82 galaxies for R=2 clusters, and 100 {\it versus} 131
galaxies for R=3 clusters. Moreover, do notice that the cut--off $S/N$
is actually very different: while we set this threshold to $4 \sigma$,
P96 report as detection limit the 95th percentile (\ie approximately
the $2 \sigma$ level) in their simulations.\\

As for the expected contamination rate, we ran our algorithm on the
EIS catalogue after having randomized galaxy positions and
magnitudes. This time, no clusters were added to the galaxy background
distribution, as we were interested in the false detection rate
associated to a given threshold $S/N$.  These simulations supply $\sim
14 $ spurious candidates per square degree if the threshold is set at
$S/N = 3$, and $\sim 1.3 $ per square degree if it is set at $S/N =
4$.  P96 report an estimated contamination rate for
their final catalogue (of $79$ candidates) of at most $30 \%$ in their
$5.1$ square degree area, corresponding to a surface density of
contaminants of $\sim 5 $ per square degree for PDCS. However, a
direct comparison between the two results is very difficult since the
data are different and the algorithms have very different selection
methods.

\section{The catalogues}\label{catalogos}

Our final catalogues contain the following entries for each one of the
candidate clusters found in EIS-wide patches A and B: right ascension and declination
(equinox 2000), indicative angular size $2.5 \sigma_{ang}$ (in
arcmin), combined total probability expressed as a $S/N$, $m^*$,
redshift and richness ($N_{0.5}$) estimates.\\

\subsection{Defining our data set}

ESO provides on the web the single frame catalogues for the even and
the odd coverage of the EIS patches
(http://www.eso.org/science/eis/). We thus had to assemble these data
and build two galaxy catalogues (henceforth the even and the odd) for
each entire patch, that were the basis for the cluster searches.

After careful inspection of the EIS data characteristics, we decided
to adopt the following parameters to define the galaxy catalogue on
which to perform our automated search.  The star/galaxy separation
line was drawn at $0.9$: only the objects with stellarity index lower
than this value were kept, irrespective of their magnitude.  Let us
notice that this criterion slightly differs from that adopted by Olsen
\etal (\cite{olsena}, \cite{olsenb}), where all the objects at magnitude fainter than 21
were kept, irrespective of their stellar index, introducing a sharp
step at I=21 in the galaxy number counts.\\

We excluded objects with $f > 4$ and nflag/npix $ > 0.01$ (where $f$
are the SExtractor flags, npix the number of pixels above the analysis
threshold and nflag the number of pixels flagged by WeightWatcher -
see Nonino \etal \cite{noni} for details).  The layout of the EIS mosaic is
such that adjacent odd/even frames have a small overlap at the edges,
especially in declination and we further did cure double detections in
frame borders (\ie for pixel values $X \le 100$, $X \ge 1950$, $Y \le
300$ or $Y \ge 1600$) using a $2$ arcsec separation cut--off.

We limited the data set to $18 \le I \le 22.0$, aiming at guaranteeing
a high level of completeness of the data. We note that Nonino \etal
(\cite{noni}) estimate that the single-frame odd and even catalogues are $80
\%$ complete to $I = 23$ for a typical EIS frame.\\

Finally, and for technical reasons, we trimmed each patch so as to
have clear-cut rectangular edges. This translates into the limits
given in table \ref{tab:patchlimits}, and in the following number of
galaxies: 57366 for patch A--even, 57553 for patch A--odd, 25456 for
patch B--even and 25445 for patch B--odd. In summary, we have four
catalogues, two (the even and the odd) for each patch, where we shall
independently run our algorithm.

\begin{table}[htbp]
\centering
\begin{tabular}{|c|c|c|c|c|}
\hline \hline
 patch &  $\alpha_{min}$ &  $\alpha_{max}$ &  $\delta_{min}$ &  $\delta_{max}$  \\
       & $^{(h}$ $^m$ $^{s)}$ & $^{(h}$ $^m$ $^{s)}$ &  $^{(o}$ ' ''$^)$ & $^{(o}$ ' ''$^)$ \\
\hline
A & 22 35 31 & 22 49 41 & -40 28 37  & -39 27 36  \\
\hline
B & 00 44 30  & 00 54 01 & -29 53 24 & -29 17 24  \\
\hline \hline
\end{tabular}
\caption{Limits of the regions to be analyzed by our algorithm 
(coordinates are expressed in equinox J2000).}
\label{tab:patchlimits}
\end{table}

\subsection{Defining our $S/N$ thresholds}\label{thresholds}

The choice of a threshold value in $S/N$ for cluster detection has to be
a compromise between the will to achieve a high level of
completeness and the need to avoid unacceptable high values of
contamination.

In section \ref{overallcomplete} we have shown that a cut--off value
of $S/N = 4 \sigma$ gives a number of contaminants of $\sim 1.3 $ per
square degree, while guaranteeing a high level of completeness ($\ge
97\%$ for all three richness classes $R = 1 - 3$ up to a redshift of
$0.8$ - see figure \ref{fig:completeness}).

If we had in our hands two good quality even and odd catalogues for
each patch, we could just apply such cut--off on both catalogues and
get two final lists of cluster candidates that should differ by a
small number, caused by false detections in regions where \eg
SExtractor detected spurious objects in the spikes of bright saturated
stars or satellite tracks.

Unluckily, this does not appear to be the case for the data,
especially in what concerns patch A, whose CCD frame quality is quite
variable (Olsen \etal \cite{olsenb}).  As a result, applying a blind cut--off
of $4 \sigma$ to both catalogues would be too restrictive a choice,
and therefore we decided to follow a more flexible approach.

For each patch we first built, for the corresponding even and odd galaxy 
catalogue, a catalogue of cluster candidates with a $S/N$ cut--off of 3. Then, 
from these catalogues we selected:

\begin{itemize}
\item[] \underline{{\it Class 1} candidates:} present in both even and odd 
catalogues (which implicitly means that they have $S/N \ge 3.0 \sigma$ in both 
catalogues) with $S/N \ge 4.0 \sigma$ in at least one of them;\\
\item[] \underline{{\it Class 2} candidates:} present in one catalogue only 
but having $S/N \ge 4.0 \sigma$. 
\end{itemize}

In this way, and for each patch, we do not reject those candidates
that are present only in one catalogue; in fact, very often, these
single detections include, other than flagrant false candidates,
systems non detected in one of the catalogues due to shallower images,
worse seeing conditions or general lower photometric quality.

In terms of contamination, our choice is equivalent to applying a $S/N \ge 4.0$ cut--off to the even/odd catalogues separately and the same
holds for completeness, but with the further constraint for {\it class
1} candidates of being present also in the odd/even catalogue with $S/N \ge 3.0$ (no further constraint for {\it class 2} candidates).

\subsection{Cluster candidates}\label{cands}

\subsubsection{Patch A}

{\it Class 1} candidates in patch A total $41$ (see table 
\ref{tab:CATFINAL_Amatched}), which translates into a
projected number density $\Sigma \sim 13$ clusters per square
degree (this patch covers $\sim 3$ square degrees).  Three further
{\it class 1} candidates were detected in patch A but are not listed in
Table \ref{tab:CATFINAL_Amatched}. Two were cases of a bright star
split into many objects and causing false detections, and a third one
was a nearby low surface brightness galaxy again split into many
components. 
The total expected number of spurious detections in this area (see
section \ref{overallcomplete}), is $< 4$. 

Table \ref{tab:CATFINAL_Amatched} lists, for each candidate, its right ascension,
declination (both at the year 2000), a typical size ($2.5\sigma_{ang}$, where
$\sigma_{ang}$ is the size of the spatial detection filter in arcmin),
the $S/N$ in the even and odd galaxy catalogues, the output $m^*$, the
estimated cluster redshift and the richness indicator $N_{0.5}$
(Bahcall, \cite{bahcall81}). The last column reports identifications with other
catalogues, whenever is the case.

The indicated size is the mean value of what was measured from the
even and odd catalogues respectively. The same was computed for all
the other quantities in the table ($S/N$ excluded, obviously), except
whenever at least one of the two next cases occurred: (1) The redshift
estimates from the even and odd catalogues differed in absolute value
by more than 0.1. This happened for a small
percentage of the entries (8 over 41 \ie 20 \%), suggesting that our
{\it 'a posteriori'} redshift estimate is quite robust ``internally''
(between even and odd catalogues).  (2) The richness estimate was
hampered because the candidate was attributed $m_3 + 2 > 22$ (the
adopted magnitude limit) in the even catalogue (noted $(e)$) or in the
odd (marked $(o)$) or in both ($(eo)$).
In these cases, results issued from both even and odd catalogues were listed.

Furthermore, candidates marked with an asterisk have only a lower
limit for the $S/N$, obtained using the spatial information alone.  For
these candidates we preferred to avoid quoting the total $S/N$ as their
$P_{lum}$ was extremely low, possibly an artifact due to bad data
quality: a careful inspection revealed that local faint galaxy counts 
were significantly below those of patch A as a whole thus causing, as 
a consequence, an overestimate of the significance of the bright galaxy 
excess.\\

\begin{table*}[htbp]
\centering
\begin{tabular}{|c|c|c|c|c|c|c|c|c|}
\hline \hline
 $\alpha_{2000}$ & $\delta_{2000}$ & 2.5$\sigma_{ang}$ & $S/N_{even}$ & 
 $S/N_{odd}$ & $m^*$ & $z_{est}$ & $N_{0.5}$ & Comments \\
\hline
22 35 44.4   &  -39 35 52.2   &   1.33   &   3.85   &   4.08   & 19.35 &  0.55   &  20  & \\
22 35 59.4   &  -39 35 53.8   &   1.07   &   5.42   &   $\ge$ 4.34 *  & 18.00 &  0.30   &  15  & EIS 2236-3935 \\
22 36 01.6   &  -39 46 24.3   &   2.50   &   5.19   &   5.15   & 21.75 &  1.05   &  17  & \\
22 36 17.9   &  -40 17 39.3   &   2.50   &   6.24   &   6.32   & 20.15 &  0.70   &  26  & EIS 2236-4017 \\
22 38 03.8   &  -39 34 02.1   &   3.54   &   4.16   &   4.06   & 18.45 &  0.40   &  11  & EIS 2238-3934 \\
22 38 33.4   &  -40 01 20.8   &   1.77   &   5.10   &   4.29   & 20.25 &  0.75   &  26  & EIS 2238-4001 \\ 
22 38 45.9   &  -39 53 18.7   &   1.25   &   4.36   &   4.57   & 19.80/19.90 &  0.6/0.7   &  21/24$(eo)$  & EIS 2238-3953 \\
22 38 49.3   &  -39 34 16.8   &   2.50   &   3.72   &   4.04   & 22.00/22.00 &  1.1/1.1   &  11/16$(eo)$  & \\
22 38 55.1   &  -40 23 46.1   &   1.07   &   4.50   &   3.20   & 19.85 &  0.65   &  16  & \\
22 39 16.7   &  -40 20 13.1   &   1.25   &   4.03   &   4.24   & 18.45 &  0.40   &  18  & \\
22 39 34.2   &  -39 37 22.2   &   1.07   &   3.57   &   4.22   & 20.60/20.70 &  0.8/0.8   &  19/21$(eo)$  & \\ 
22 40 16.5   &  -40 25 33.9   &   0.88   &   4.74   &   4.95   & 19.70/19.30 &  0.6/0.5   &  28/23$(e)$  & \\
22 40 58.2   &  -40 06 12.5   &   3.54   &   5.14   &   4.70   & 20.80/21.10 &  0.9/0.9   &  35/27$(e)$  & \\
22 41 16.6   &  -39 41 00.5   &   3.02   &   3.82   &   5.56   & 19.80/22.00 &  0.6/1.1   &  27/22  & \\
22 41 19.7   &  -40 00 44.1   &   2.13   &   5.00   &   5.81   & 21.50/21.60 &  1.0/1.0   &  22/25$(eo)$  & EIS 2241-4001 \\ 
22 41 33.7   &  -40 20 58.1   &   1.25   &   3.67   &   4.30   & 18.45 &  0.40   &  14  & \\
22 41 41.6   &  -39 49 56.7   &   3.54   &   5.91   &   6.86   &  18.50 &  0.40   &  19  & EIS 2241-3949\\
22 43 17.4   &  -39 51 54.3   &   2.50   &   6.05   &   7.50   &  19.30 &  0.55   &  29  & ACO1055/EIS 2243-3952 \\ 
22 43 24.3   &  -40 25 32.8   &   2.13   &   5.27   &   5.94   & 19.00 &  0.50   &  27  & EIS 2243-4025 \\
22 43 36.1   &  -40 00 42.9   &   1.07   &   4.57   &   $\ge$ 3.93 *  & 18.00 &  0.30   &  14  & \\
22 43 42.4   &  -39 41 48.8   &   3.54   &   3.91   &   4.50   &  18.80 &  0.45   &  16  & \\ 
22 44 02.5   &  -40 05 19.5   &   0.88   &   4.87   &   4.76   & 20.45 &  0.75   &  20  & \\ 
22 44 13.6   &  -39 37 05.0   &   1.51   &   3.35   &   4.04   &  20.10/19.10 &  0.7/0.5   &  19/18$(e)$  & \\ 
22 44 19.4   &  -39 51 43.0   &   1.51   &   3.63   &   4.11   &  19.80/19.60 &  0.6/0.6   &  14/16$(e)$  & \\
22 44 21.4   &  -40 08 09.2   &   0.88   &   4.08   &   4.20   &  18.70 &  0.40   &  11  & \\
22 44 40.4   &  -39 45 29.4   &   3.02   &   5.67   &   4.75   & 18.80/18.90 &  0.4/0.5   &  18/21$(e)$  & \\
22 44 58.5   &  -40 26 35.9  &   1.07   &   3.88   &   5.47   & 19.70/18.50 &  0.6/0.4   &  16/17$(e)$  & \\
22 45 13.2   &  -39 54 09.9   &   0.88   &   4.88   &   5.07   & 19.50/19.40 &  0.6/0.6   &  12/10$(e)$  & \\ 
22 45 34.9   &  -40 13 39.1   &   1.07   &   3.40   &   4.53   & 19.90/21.80 &  0.7/1.1   &  14/18  & \\
22 45 48.1   &  -39 33 58.1   &   1.25   &   5.04   &   4.10   & 18.20/18.50 &  0.4/0.4   &  16/13$(o)$  & \\ 
22 46 01.8   &  -39 53 50.0   &   3.54   &   4.64   &   4.00   &    22.00/22.00 &  1.1/1.1   &   7/6$(o)$  & \\
22 46 12.4   &  -39 53 14.2   &   1.07   &   3.69   &   4.41   & 19.30/20.60 &  0.5/0.8   &  16/18  & \\
22 46 26.9   &  -40 04 15.8   &   0.88   &   4.70   &   4.62   & 21.20/20.30 &  0.9/0.7   &  11/12$(eo)$  & \\
22 46 48.8   &  -40 12 41.9   &   2.39   &   4.05   &   4.31   &  18.90 &  0.45   &  13  & EIS 2246-4012 \\
22 47 53.2   &  -39 35 46.8   &   1.51   &   4.26   &   4.22   & 18.60/19.40 &  0.4/0.6   &  14/14  & \\ 
22 47 54.2   &  -39 46 33.7   &   1.88   &   5.16   &   4.12   & 22.00/19.20 &  1.1/0.5   &  12/16  & LP96 \\
22 48 27.9   &  -39 50 48.2   &   3.02   &   4.40   &   4.45   & 19.35 &  0.55   &  19  & EIS 2248-3951 \\
22 49 04.0   &  -39 42 36.5   &   1.33   &   4.05   &   3.04   &  19.20 &  0.55   &  16  & \\ 
22 49 15.7   &  -39 37 32.6   &   1.07   &   4.39   &   3.77   &  21.20/20.60 &  0.9/0.8   &  18/16$(eo)$  & \\ 
22 49 32.1   &  -39 58 04.6   &   0.88   &   4.10   &   4.11   & 20.40/20.40 &  0.8/0.8   &  13/16$(eo)$  & EIS 2249-3958 \\ 
22 49 32.7   &  -40 16 31.0   &   1.69   &   3.15   &   5.19   & 19.80/19.90 &  0.6/0.7   &  24/29$(eo)$  & EIS 2249-4016 \\
\hline \hline
\end{tabular}
\caption{{\it Class 1} cluster candidates of patch A, ordered in right ascension. 
Coordinates ($\alpha$,$\delta$)$_{2000}$ and approximate angular size 
2.5$\sigma_{ang}$ (expressed in arcmin) are the mean values of the quantities 
issued from the even and the odd catalogues. We report values of $m^*$, $z$
and $N_{0.5}$ for both the even and the odd catalogue if the
discrepancy amongst at least one of the first two quantities was large
or if $m_3 + 2 > 22$ (22 being our adopted magnitude limit) 
in the even catalogue, signalled by $(e)$, or in the odd one, $(o)$,
or in both, $(eo)$, making the estimate of $N_{0.5}$ quite uncertain. 
Some further remarks on individual entries: we note that a bright galaxy at the center in the line of
sight of candidate $22^h 38^m 49.3^s, -39^o 34' 16.8''$ may have caused some
($4$ at the most) spurious detections, but it is hard to say as the
background noise is enhanced in that area. Candidate $22~43~17.4,
-39~51~54.3$ is close to a large foreground spiral (ESO 345-G046) 
which arms were broken into pieces by the SExtractor procedure, 
thus producing spurious detections 
(see text). Finally, candidate $22^h 44^m 40.4^s, -39^o 45' 29.4''$ is near a 
bright star, but galaxy detections are not affected by it (maybe 
$4$ of the objects detected in the cluster periphery are doubtful).
}
\label{tab:CATFINAL_Amatched}
\end{table*}

As for {\it class 2} candidates in this patch, table
\ref{tab:CATFINAL_Aunmatched_secondtry} lists $29$ detections: those
that survived a visual scrutiny aimed at rejecting obvious cases of
false signals (satellite trails, bright star refraction spikes and
other artifacts). For each cluster candidate the same quantities as in Table
\ref{tab:CATFINAL_Amatched} are listed.

\begin{table*}[htbp]
\centering
\begin{tabular}{|c|c|c|c|c|c|c|c|c|}
\hline \hline
 $\alpha_{2000}$ & $\delta_{2000}$ & 2.5$\sigma_{ang}$ & $S/N_{even}$ & 
 $S/N_{odd}$ & $m^*$ & $z_{est}$ & $N_{0.5}$ & Comments \\
\hline
22 36 55.8   &   -40 24 11.4   &   1.00   &    -     &   4.08   & 19.90 &   0.7    &   15  & \\
22 37 04.5   &   -39 59 42.4   &   1.41   &    -     &   4.02   & 18.70 &   0.4    &   16  & \\
22 37 08.5   &   -40 00 22.2   &   0.35   &    -     &   4.50   & 18.00 &   0.3    &   21  & EIS 2237-4000 \\
22 37 47.2   &   -40 06 14.7   &   1.00   &    4.39  &    -     & 21.90 &   1.1    &   15  & \\
22 38 38.7   &   -40 00 03.6   &   1.41   &    -     &   5.32   & 21.30 &   1.0    &   20$(o)$  & \\
22 38 41.8   &   -40 12 10.8   &   1.41   &    4.10  &    -     & 21.20 &   0.9    &   12$(e)$  & \\
22 39 01.6   &   -39 47 14.7   &   1.00   &    4.08  &    -     & 19.30 &   0.5    &   18  & \\
22 40 06.4   &   -40 20 02.1   &   0.71   &    -     &   4.39   & 18.40 &   0.4    &   21  & \\
22 40 07.4   &   -39 51 27.2   &   0.35   &    4.28  &    -     & 21.50 &   1.0    &   11$(e)$  & \\
22 40 08.5   &   -40 21 01.1   &   1.41   &    5.53  &    -     & 18.70 &   0.4    &   16  & EIS 2240-4021 \\
22 40 36.2   &   -39 32 31.6   &   0.35   &    -     &   4.15   & 22.00 &   1.1    &   10$(o)$  & \\
22 40 38.5   &   -39 57 27.1   &   0.35   &    -     &   4.24   & 21.00 &   0.9    &    8$(o)$  & \\
22 40 51.6   &   -39 40 58.1   &   0.71   &    -     &   4.01   & 21.70 &   1.1    &   19$(o)$  & \\
22 41 11.2   &   -40 21 22.3   &   1.41   &    4.18  &    -     & 18.90 &   0.5    &   11  & \\
22 42 39.2   &   -39 45 56.4   &   1.00   &    -     &   5.27   & 22.00 &   1.1    &   19$(o)$  & \\
22 42 59.8   &   -39 32 21.7   &   0.35   &    4.15  &    -     & 18.60 &   0.4    &   18  & \\ 
22 43 04.6   &   -40 15 37.0   &   0.71   &    -     &   4.47   & 20.40 &   0.8    &   29$(o)$  & \\
22 43 17.1   &   -39 39 33.9   &   0.71   &    4.20  &    -     & 18.80 &   0.4    &   22  & \\ 
22 44 53.3   &   -39 37 39.2   &   0.35   &    -     & $\ge$ 4.66 *  & 18.00 &   0.3    &   27  & \\ 
22 45 33.1   &   -39 39 52.7   &   0.50   &    -     & $\ge$ 4.24 *  & 18.00 &   0.3    &   35  & \\ 
22 45 58.5   &   -39 31 04.8   &   0.35   &    4.07  &    -     & 21.20 &   0.9    &    7$(e)$  & \\ 
22 45 58.6   &   -39 43 21.3   &   0.71   &    -     & $\ge$ 4.69 *  & 18.10 &   0.3    &   29  & \\ 
22 46 53.0   &   -39 41 03.4   &   0.35   &    -     &   4.07   & 21.70 &   1.1    &    7$(o)$  & \\ 
22 47 43.2   &   -39 55 29.7   &   1.00   &    4.33  &    -     & 21.10 &   0.9    &    8  & \\
22 47 52.1   &   -40 02 05.6   &   0.71   &    -     &   4.03   & 19.30 &   0.5    &   15  & \\
22 48 55.3   &   -40 15 37.0   &   0.71   &    -     &   4.02   & 18.00 &   0.3    &   13  & EIS 2248-4015 \\
22 48 56.7   &   -39 32 32.3   &   0.35   &    4.52  &    -     & 19.70 &   0.6    &   20  & \\ 
22 48 57.6   &   -40 22 44.7   &   1.00   &    4.59  &    -     & 20.50 &   0.8    &   26  & \\
22 49 29.2   &   -40 05 27.7   &   0.35   &    4.17  &    -     & 22.00 &   1.1    &   12$(e)$  & \\
\hline \hline
\end{tabular}
\caption{{\it Class 2} patch A cluster candidates. Columns and notes follow the scheme 
of the ones in table \ref{tab:CATFINAL_Amatched}.} 
\label{tab:CATFINAL_Aunmatched_secondtry}
\end{table*} 

The reason for a candidate - not being an obvious artifact of the data
- to be present only in one of the two catalogues of patch A has to be
related to the worse quality of the corresponding CCD frames in the
catalogue where it is missing. The variables at work are numerous
(such as seeing conditions, background noise, etc.) and not always
easy to define. Often, a noisy background results in a higher number
of - possibly spurious - galaxy detections by SExtractor, while a
worse seeing invariably results in a lower number of galaxy
detections.

The large number of {\it class 2} detections is an indication of the
large range in quality of the CCD data available for patch A. In fact,
if we check the ratio of {\it class 2} to {\it class 1} objects, we
get $29/41$ or, in other words, {\it class 2} candidates represent a
percentage of $\sim 41 \%$ of the total sample. Olsen \etal report,
for the same ratio, $15/20$, which translates into $\sim 43 \%$ of
their whole sample.

\begin{figure*}[htbp]
\centering
\psfig{figure=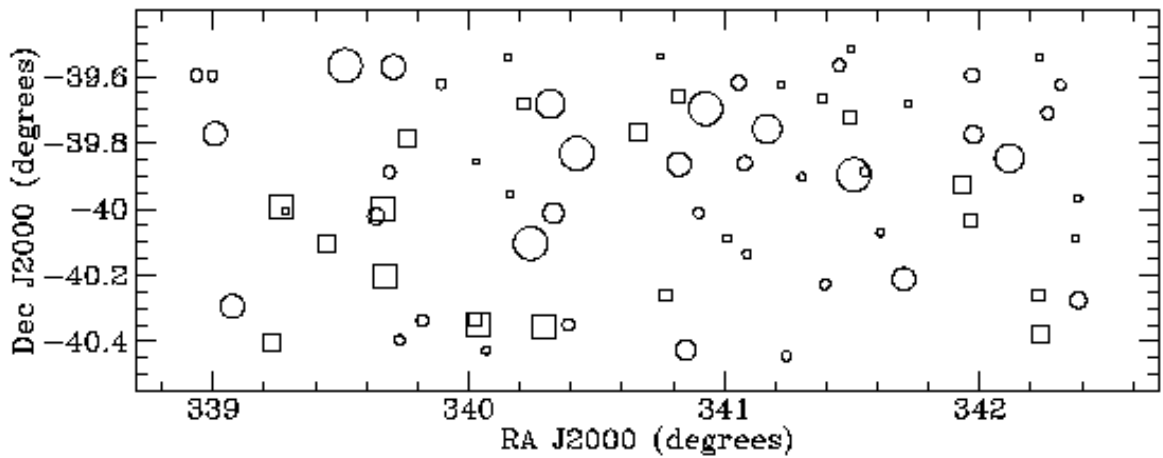,width=16cm} 
\caption{Map of patch A candidates. Circles indicate {\it class 1} systems 
while squares designate {\it class 2} candidates. Coordinates and sizes 
(proportional to $\sigma_{ang}$) are mean values from even and odd catalogues. 
Do notice that some {\it class 2} candidates (namely at, roughly, 
($339.3^o$, $-40^o$) and ($340^o$, $-40.35^o$)) may misleadingly appear
to have escaped our double detection elimination procedure but the
reason is that apparently superimposed detections are actually unmatched 
({\it class 2}) ones flagged in different catalogues (even or odd).}
\label{fig:mapdets_A}
\end{figure*}

Figure \ref{fig:mapdets_A} shows the sky distribution of our {\it class 1} 
and {\it class 2} candidates, circles and squares respectively.

Adding {\it class 1} to {\it class 2} candidates provides a total of
$70$ cluster candidates for patch A, corresponding to a surface
density of $23$ per square degree, with $1.3$ expected spurious
detections per square degree. Out of this total, 40\% can be
identified as richness class $0$ objects and the same percentage
encloses $R \sim 1$ systems, while the remaining 20\% seem to be
richness $R \sim 2$ clusters - see the upper panel of figure
\ref{fig:N05hist_AandB}.  The corresponding redshift distribution for
the total Patch A sample is plotted in the upper part of figure
\ref{fig:zhist_AandB}: the distribution of {\it class 1} candidates
does not seem to differ significantly from that of {\it class 2} candidates, 
both covering the redshift range $0.3$ to $1.0$ (the excess tail present
for {\it class 2} candidates in the last redshift bin may be an
artifact, due to those candidates whose $m^*$ approaches (or would exceed) 
$I = 22$, the limiting magnitude of our galaxy catalogue).

A detailed comparison of our catalogue of cluster candidates with that
of Olsen \etal (\cite{olsenb}) will be presented in section \ref{compolsen}.
Here, let us notice that in patch A there are four cluster candidates
detected by previous independent works and listed in NED. Abell
cluster $ACO1055$, that was selected as a {\it class 1} candidate by
us with a high $S/N$.  Cl2245-4002 (LP96) of Lidman $\&$ Peterson (\cite{lid})
is well identified with our entry $22^h 47^m 54.2^s, -39^o 46'
33.7''$.  Two EDCC cluster candidates ($EDCC 163$ and $EDCC 169$;
Lumsden \etal \cite{lum})
do not turn up in our sample. We decided to point our algorithm
directly at the NED coordinates of these two clusters while using the
galaxy catalogue at $I_{lim} = 23$, just to check if their absence
from our list could be due to the magnitude cut--off chosen. This
approach managed to retrieve $EDCC 169$ with $S/N_{even} = 3.29$ and
$S/N_{odd} = 3.84$, while $EDCC 163$ is still missing. According to
the Lumsden \etal (\cite{lum}) table, both candidates are considered to be
poor R $< 0$ systems, which may account for our results.

Out of these 4 objects, only Abell $ACO1055$ has spectroscopic
confirmation (at $z = 0.0322$) and, consequently, a reliable richness
estimate ($R = 0$) in the literature (\cf NED database).
The redshift we estimate for this system is higher
($z_{even} = 0.6$, $z_{odd} = 0.5$) and the richness we compute ($R
\simeq 2$) is, consequently, considerably incorrect.  Notice though that 
visual inspection indicates that a lot of spurious ``galaxies'' were detected
by SExtractor in the spiral arms of the foreground bright galaxy ESO
345-G046 (the effect is more serious for the even catalogue), and this
has possibly affected our results, namely the determination of $m^*$.

\subsubsection{Patch B}
	     
{\it Class 1} candidates in patch B total $21$, which translates into
$\Sigma \sim 15$ clusters per square degree (this patch covers,
approximately, 1.4 square degrees). This is a slightly higher density
than that found for patch A, possibly because of better data quality
(patch B had better observing conditions). Note that Olsen and
collaborators also find some difference: they report $14$ clusters per
square degree in patch A while the value rises to $17.2$ for patch B.
This difference between both patches is, however, negligible within
uncertainties.

Table \ref{tab:CATFINAL_Bmatched} lists, with the same format already used for 
patch A candidates, our {\it class 1} patch B cluster candidates. 

The simulations of section \ref{overallcomplete} allow us to predict 
$1.3 \times 1.4 \sim 2$ spurious detections associated to the $21$ candidates.

\begin{table*}[htbp]
\centering
\begin{tabular}{|c|c|c|c|c|c|c|c|c|}
\hline \hline
 $\alpha_{2000}$ & $\delta_{2000}$ & 2.5 $\sigma_{ang}$ & $S/N_{even}$ &
$S/N_{odd}$ & $m^*$ & $z_{est}$ & $N_{0.5}$  & Comments \\
\hline
 00 45 21.0   &  -29 23 35.5  &  3.54   & $\ge$ 6.34 *  &  $\ge$ 7.32 *  & 18.40 &  0.40  &  24   & EIS 0045-2923 \\
 00 45 40.5   &  -29 50 54.2  &  0.88   &    4.48  &    3.91    & 20.60/19.70 &  0.8/0.6   &  18/16$(eo)$   & \\
 00 45 44.0   &  -29 47 57.5  &  2.13   &    4.81  &    4.67    & 22.00/19.10 &  1.1/0.5   &  17/23   &EIS 0045-2948\\
 00 46 07.5   &  -29 51 28.7  &  0.88   &    3.99  &    4.39    & 20.05 &  0.70   &  17   & EIS 0046-2951 \\
 00 46 08.1   &  -29 23 40.8  &  1.77   &    3.90  &    4.50    & 18.95 &  0.45   &  19   & \\
 00 46 33.9   &  -29 39 03.2  &  1.69   &    4.90  &    3.95    & 21.10/20.00 &  0.9/0.7   &  22/18 $(eo)$   & \\
 00 46 35.9   &  -29 25 38.8  &  0.88   &    4.03  &    3.34    & 18.90/18.70 &  0.5/0.4   &   7/7$(e)$   & \\
 00 47 22.3   &  -29 48 33.7  &  0.88   &    4.49  &    3.72    & 21.20/20.80 &  0.9/0.9   &  19/14$(eo)$   & \\ 
 00 48 31.6   &  -29 42 06.6   &  0.88   &    4.34  &    4.34    & 20.80 &  0.85   &  15   & EIS 0048-2942 \\ 
 00 49 19.0   &  -29 22 45.2   &  0.88   &    3.21  &    4.03    & 18.95 &  0.45   &   8   & \\
 00 49 23.2   &  -29 30 43.0   &  3.02   &    5.00  &    5.56    & 18.55 &  0.40   &  19   & ACO84/EIS 0049-2931 \\
 00 49 23.3   &  -29 47 21.4   &  1.07   &    3.14  &    4.25    & 20.30/20.30 &  0.7/0.7   &  11/17$(eo)$   & \\
 00 49 39.5   &  -29 34 33.1   &  3.54   &    5.76  &    5.63    & 18.85 &  0.45   &   9   & \\
 00 50 04.5   &  -29 41 13.5   &  0.88   &    3.88  &    4.71    & 21.40/22.00 &  1.0/1.1   &  12/13$(eo)$   & EIS 0050-2941 \\ 
 00 50 09.4   &  -29 25 12.4   &  3.02   &    3.43  &    4.29    & 20.30/22.00 &  0.7/1.1   &  9/17$(eo)$   & \\
 00 51 47.6   &  -29 29 20.2   &  3.54   &    5.19  &    4.97    & 21.05 &  0.90   &   8   & \\
 00 51 49.6   &  -29 45 20.1   &  0.88   &    4.21  &    4.29    & 20.10/20.50 &  0.7/0.8   &  11/15$(o)$   & \\
 00 51 50.5   &  -29 38 05.3   &  0.88   &    4.76  &    3.36    & 21.80/19.90 &  1.1/0.7   &   13/4$(eo)$   & \\
 00 52 49.3   &  -29 27 49.5   &  1.07   &    4.83  &    4.18    & 21.80/21.90 &  1.1/1.1   &  17/15$(eo)$   & EIS 0052-2927 \\ 
 00 53 07.1   &  -29 47 40.7   &  2.65   &    4.04  &    3.43    & 21.50/20.60 &  1.0/0.8   &  11/11$(o)$   & \\
 00 53 38.1   &  -29 28 41.8   &  0.88   &    3.00  &    4.09    & 21.10/21.50 &  0.9/1.0   &  12/9$(eo)$   & \\
\hline \hline
\end{tabular}
\caption{{\it Class 1} cluster candidates of patch B, ordered in right ascension. 
Quantities listed and notes used are equivalent to those reported in table 
\ref{tab:CATFINAL_Amatched}. Candidate $00^h 51^m 50.5^s, -29^o 38' 05.3''$ is near a 
bright star (and $4$ of the objects detected in the cluster periphery are 
doubtful).} 
\label{tab:CATFINAL_Bmatched}
\end{table*}

\begin{figure*}[htbp]
\centering
\psfig{figure=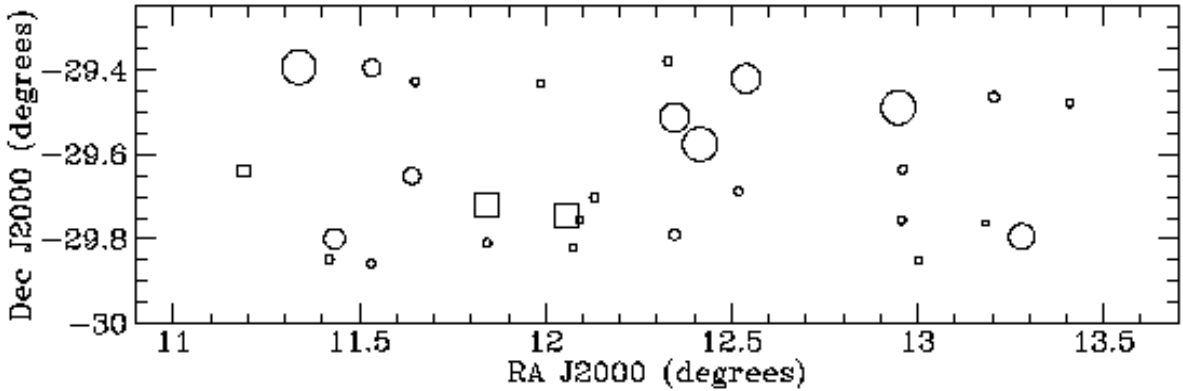,width=16cm}
\caption{Same as figure \ref{fig:mapdets_A} but for patch B candidates.}
\label{fig:mapdets_B}
\end{figure*}

Very much like for patch A, in 71 \% of the cases (15 out of 21),
the redshift estimates from the even and odd catalogues differ in
absolute value of less than $0.1$. \\

As for {\it class 2} candidates in this patch, table
\ref{tab:CATFINAL_Bunmatched_secondtry} lists all $8$ detections that
survived accurate visual inspection.

Figure \ref{fig:mapdets_B} shows the sky
distribution of our {\it class 1} and {\it class 2} candidates,
respectively.

\begin{table*}[htbp]
\centering
\begin{tabular}{|c|c|c|c|c|c|c|c|c|}
\hline \hline
 $\alpha_{2000}$ & $\delta_{2000}$ & 2.5 $\sigma_{ang}$ & $S/N_{even}$ &
$S/N_{odd}$ & $m^*$ & $z_{est}$ & $N_{0.5}$ & Comments\\
\hline
 00 44 45.0  &  -29 38 17.1  &  0.71  &   4.18  &    -    & 21.30 &   1.0   &  20$(e)$  &\\
 00 47 21.9  &  -29 43 16.9  &  1.41  &   -     &   4.90  & 19.70 &   0.6   &  11      & \\
 00 47 56.7  &  -29 26 00.1  &  0.35  &   -     &   4.68  & 22.00 &   1.1   &   9$(o)$  &\\
 00 48 13.5  &  -29 44 44.3  &  1.41  & $\ge$ 4.21 *  &    -      & 18.00   &   0.3   &  12  &\\
 00 48 18.2  &  -29 49 09.5  &  0.35  &   -     &   4.27  & 20.10 &   0.7   &  11  & \\
 00 48 22.1  &  -29 45 21.4  &  0.35  &   5.25  &    -    & 18.00 &   0.3   &  18  & \\
 00 52 00.5  &  -29 50 58.3  &  0.35  &   -     &   4.00  & 21.80 &   1.1   &  11$(o)$  & \\
 00 52 43.6  &  -29 45 45.3  &  0.35  &   4.18  &    -    & 21.70 &   1.1   &  14$(e)$  & \\
\hline \hline
\end{tabular}
\caption{{\it Class 2} patch B cluster candidates. Columns and notes are the 
same that were used in table \ref{tab:CATFINAL_Bmatched}.} 
\label{tab:CATFINAL_Bunmatched_secondtry}
\end{table*}

Also for patch B holds the discussion already made for {\it class 2}
candidates in patch A. It is to be noticed that now the ratio of {\it
class 2} to {\it class 1} candidates is much lower than the one
obtained for patch A: $8/21$ or a percentage of $28 \%$ of the total
sample ({\it versus} the previous $41 \%$), suggesting that the
quality of the CCD data available for patch B is indeed much more
homogeneous. We note here that the same type of ratio produced with
Olsen \etal's candidates for this patch is $9/10$ or, in terms of
percentage of their whole sample, $\sim 47 \%$, which remains quite
similar to the figure they produce for patch A.

\begin{figure}[htbp]
\centering
\psfig{figure=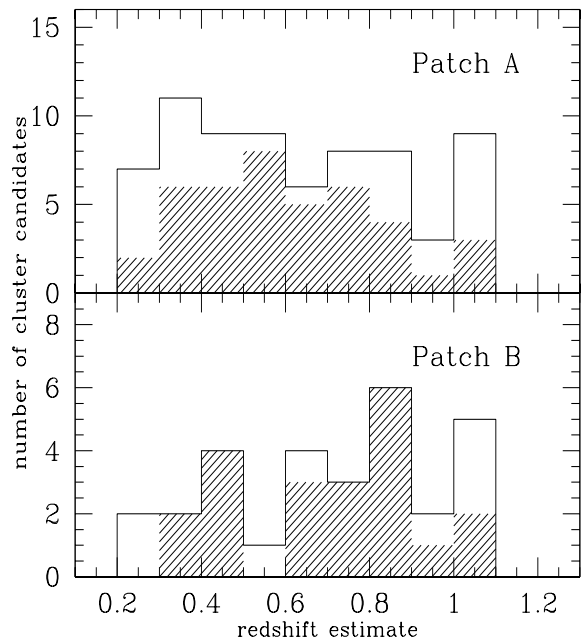,height=9.cm,width=8.cm}
\caption{Redshift distribution for the cluster candidates flagged in
patch A - upper panel - and patch B - lower panel. Hatched histograms
specify matched ($class 1$) detections among the global distribution
(empty histograms). In this figure the $z$ plotted for each cluster
candidate is the mean value obtained from even and odd catalogues. 
}
\label{fig:zhist_AandB}
\end{figure}

\begin{figure}[htbp]
\centering
\psfig{figure=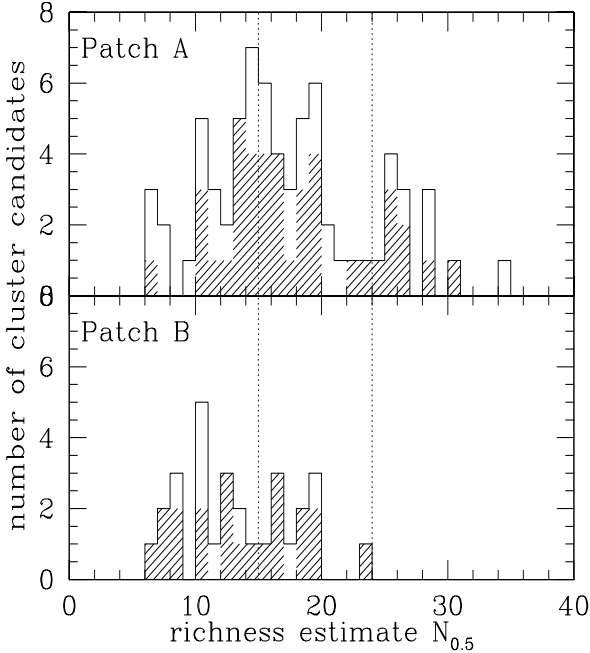,height=9.cm,width=8.cm}
\caption{Distribution of the richness estimate $N_{0.5}$ for the
cluster candidates flagged in patch A - upper panel - and patch B -
lower panel. Hatched histograms specify matched detections among the
global distribution (empty histograms).  The dotted lines mark the
approximate limits between Abell richness classes 0 and 1 (at
$N_{0.5} \sim 15$) and Abell richness classes 1 and 2 ($N_{0.5} \sim
24$).  The $N_{0.5}$ plotted for each cluster candidate 
is the mean value obtained from even and odd catalogues.  }
\label{fig:N05hist_AandB}
\end{figure}

Adding {\it class 1} to {\it class 2} candidates gives a total of
$29$ cluster candidates in patch B, corresponding to a surface density
of $\sim 21$ per square degree, with, again, the already mentioned
$1.3$ expected spurious detections per square degree. Out of these
$29$ systems, 62 \% have estimated richness class $R \sim 0$, and 38 \% 
have $R \sim 1$, as illustrated in the lower panel of figure
\ref{fig:N05hist_AandB}.  Notice that, the area probed here being
smaller than the one of Patch A, at the same limiting magnitude this
translates into a smaller sampled volume. So it is expected that the
median typical richness of the detected systems should move to lower 
values. 
The corresponding total redshift distribution is plotted in the bottom
of figure \ref{fig:zhist_AandB}. The median of the redshift
distribution for {\it class 1} candidates, patch A and B considered
together, is $ z = 0.65$, similar to that of Olsen \etal despite
having chosen a brighter cut--off in magnitude for the galaxy
catalogue ($22$ {\it versus} their $23$), while it is slightly deeper
than that of the P96 sample PDCS ($z = 0.4$).\\

There is only one cluster candidate in patch B detected by previous
independent works and listed in NED. It is cluster $ACO84$, that was
selected as a {\it class 1} candidate by us.  We estimate $0.40$ for
the redshift of this candidate (and a corresponding richness class $R
\sim 1$), while spectroscopic measurements in the literature hold a
value of $z = 0.11$ (\cf the NED database)
and $R = 0$. 
Also for patch B, a more detailed comparison between our catalogue of
cluster candidates and that of Olsen \etal (\cite{olsena}) will be presented
in section \ref{compolsen}.
 
\section{Comparison with Olsen \etal 's results}\label{compolsen}

We can now do a thorough comparison with the results that Olsen \etal
(\cite{olsena}, \cite{olsenb}) obtained by applying the P96 algorithm to patches A and B of
EIS.  In this way we will also be able to assess directly the relative
efficiency of the two algorithms (P96 and ours) on the same set of
data.

We will have to do a double check: 
first which of our candidates are new with respect to the Olsen \etal lists, 
and then which of the Olsen \etal candidates are missing from our lists.

For the comparison to be fair, we have to remember three important
points: Olsen and co-workers did their search on the total galaxy
catalogue, \ie up to magnitude $I = 23$, while we preferred to limit
it to $I = 22$.  They applied a $S/N$ cut--off of 3, while we adopted a
$S/N$ cut--off of 4 in our final cluster list and, finally, they also
imposed a cut--off in richness to their candidates, while we apply no
such sort of selection.

Bearing this in mind, we proceed with the comparisons.\\

Regarding patch A, in our {\it class 1} list we have 13 candidates in
common with Olsen \etal (\cite{olsenb}), while in the {\it class 2} list we
have only 3 candidates in common. Out of our remaining candidates, 6
from {\it class 1} and 9 from {\it class 2} are in an area of patch A
that they decided to cut--out from their search (the upper right
corner of their figure 2) because of {\it ``obvious incompleteness''
(sic)}. We did search also in this area, since our algorithm, using local
background values, is better suited for dealing with inhomogeneous
data. We therefore are left with 22 {\it class 1} candidates and 17
{\it class 2} candidates present in our list and missing in
theirs. Some of these missing candidates might have been discarded by
Olsen \etal (\cite{olsenb}) because of the richness cut--off they imposed, but
it should be noticed that amongst them there are also relatively rich
systems: 5 with estimated richness $R \sim 2$ and 25 with richness $R
\sim 1$ (conservatively counting only our $N_{0.5}$ secure values),
whose absence is thus not easy to justify.\\

In what concerns patch B, we only have 7 candidates in common with 
Olsen \etal (\cite{olsena}), all in our {\it class 1} list. 
We are therefore left with 14 {\it class 1} candidates and all 8 
{\it class 2} candidates present in our catalogues and missing in theirs.\\

On the other hand, there are candidates present in the Olsen \etal
lists that we failed to select, and we wanted to check if our missing
them could be related to the different magnitude cut--off of the
galaxy catalogue used for our search or to the different $S/N$ cut--off
adopted.

To investigate this we decided to check which would have been the $S/N$
obtained for these missing candidates by our algorithm on the galaxy
catalogue limited to their deeper magnitude cut--off ($I = 23.0$). To do
this we simply targeted directly the search to the positions of each
one of their candidates.

By doing so, we were able to retrieve all their patch A candidates
with a $S/N \ge 3$ in at least the even or the odd catalogue (the same
selection criterion they adopted), with the only exception of four of
them. One of those was commented by Olsen \etal as being a {\it
``doubtful case based on the visual inspection of the coadded image''}
and to another one our algorithm could only attribute a lower limit
$S/N$ (from the spatial part).

In patch B, with the same procedure, we assigned all their candidates
a $S/N \ge 3 \sigma$ except for three (one of them being also a case
of lower limit $S/N$ obtained from the spatial filtering).

\begin{figure}[htbp]
\epsfysize=9cm
\epsfbox{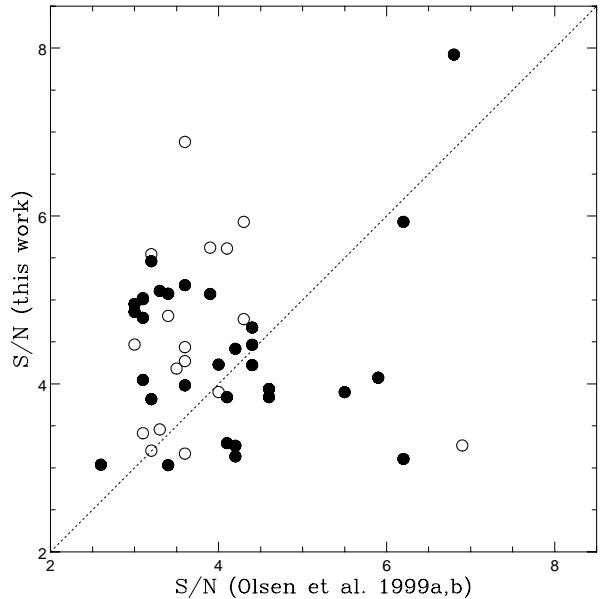}
\caption{$S/N$ obtained with our algorithm for the Olsen \etal
(1999a, 1999b) candidates {\it versus} their own $S/N$. Filled symbols stand
for patch A objects while patch B candidates are noted by empty
symbols. The $S/N_{Olsen} = S/N_{us}$ line is also marked.}
\label{fig:SNtargetsolsenAB}
\end{figure}

The plot of our $S/N$ against theirs, for all the candidates we
retrieved (common candidates), is shown in figure \ref{fig:SNtargetsolsenAB}, 
both for patch A (filled symbols) and patch B (empty symbols) objects.

One can notice that, for the majority of the cases, our $S/N$ is
greater than that obtained by Olsen \etal as would be expected: by
decoupling the spatial and the luminosity parts of the algorithm, a
candidate can be retrieved even when it does not flag a maximum
likelihood at the very same redshift value simultaneously for both
distributions, a situation that would lower its global likelihood when
using the P96 algorithm. \\

\begin{figure}[htbp]
\epsfysize=9cm
\epsfbox{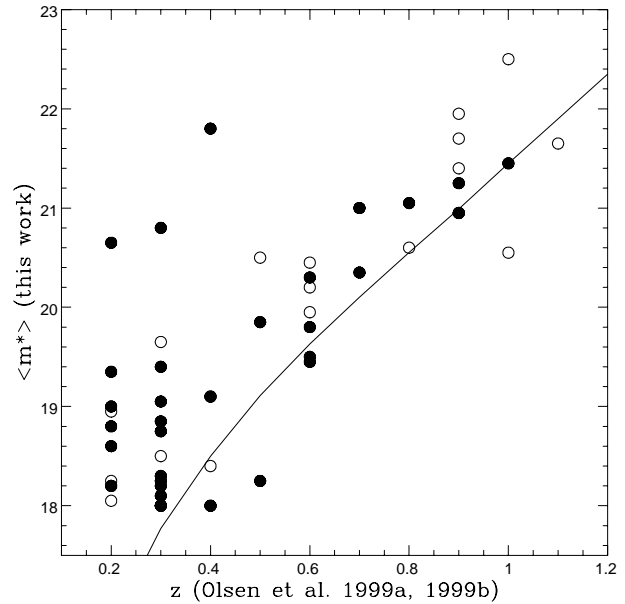}
\caption{Checking the possibility of correlation between our $m^*$ 
(average value from even and odd catalogues) and Olsen \etal 's 
redshift estimate for their candidates in patch A (filled symbols) 
and in patch B (empty symbols). 
The line is the relation deduced from Colless (1989).}
\label{fig:ms_z_23}
\end{figure}

Another interesting plot is the one showing, for the common candidates, 
our $m^*$  estimate {\it versus} the redshift estimate by Olsen \etal 
(\cite{olsena}, \cite{olsenb}). 
Figure \ref{fig:ms_z_23} plots the mean value (from the even and the
odd catalogues) of $m^*$ produced by our algorithm against the redshift
estimated by Olsen \etal for their candidates in patch A (filled circles) 
and in patch B (empty circles).  The line indicates the
redshifted value of the Colless (\cite{colless}) local mean $m^*$, affected by
k--correction typical of ellipticals (as determined by Fioc $\&$
Rocca-Volmerange \cite{fioc}).  The general distribution of the data points
does seem to follow the curve's trend. The larger redshift
systematically predicted by the Colless relationship we used is 
just a byproduct of the brighter $M^*$ estimated by Colless 
($I^* \simeq -22.2$ for $H_0 = 100$ km~s$^{-1}$Mpc$^{-1}$) {\it versus} 
the one used in P96 and by Olsen \etal ($-21.28$, also for 
$H_0 = 100$ km~s$^{-1}$Mpc$^{-1}$). This fact may also be contributing 
to the discrepancies between our estimates and the literature measures 
in the case of the Abell clusters mentioned in section \ref{cands}.

\section{Summary and Conclusions}\label{fim}

In this paper we presented a new algorithm for cluster detection that
improves on the P96 one, mainly by avoiding the need to assume a
typical physical size or a typical $M^*$ for galaxy clusters.  These
two parameters intervene in our algorithm only as typical angular
scales and a typical apparent magnitude $m^*$, bearing no ties to
fixed physical scales nor to absolute magnitudes through redshift
dependence (and chosen cosmological model).  One further advantage of
our algorithm with respect to P96's consists in the local estimate of
the background for each cluster candidate, particularly useful in the
spatial part where local inhomogeneities (due to varying conditions
during data acquisition) may hamper cluster detection in shallower
regions of the catalogue.\\

We applied this new algorithm to the EIS-wide database and did a thorough
comparison with the results obtained using P96 on the same kind of 
data. As the EIS observing conditions varied considerably
throughout the granted nights, there was a considerable spread in the
data-quality of different EIS frames (Nonino \etal \cite{noni}).  Both the
particular features of the algorithm presented in this paper, that
somehow compensate for the lack of homogeneity of the data, and the
conservative limiting magnitude adopted for cluster search try to
minimize the problems inherent to the EIS data, allowing to achieve a
higher completeness level than that obtained using P96 on the
same data. \\

In fact, we notice that the distribution of our patch A candidates
seems more homogeneous than the one presented by Olsen \etal (\cite{olsenb}):
compare our figure \ref{fig:mapdets_A} with their figure $4$, where 
they have a region around $\alpha^o \sim 341.5$, $\delta^o \sim -40$
devoid of detections, not to mention the fact that they intentionally
left out a part of the patch when running their algorithm to
``overcome'' data problems related to the lack of homogeneity. \\   

Regarding the cluster candidate surface density, a precise comparison
with the values presented in the literature is not trivial, due to
different magnitude limits adopted for the search and to the fact that
we did not apply any {\it a priori} selection on cluster richness (as
P96 for example). Just as a qualitative comparison, let us remember
that the PDCS group (P96) reports the detection of $79$ cluster
candidates over an area of $5.1$ square degrees till $I = 22.5$, or a
number density slightly above $15$. This value is consistent with the
observed X--ray $logN - logS$ around fluxes $> 6 \time 10^{-15}$ erg
cm$^{-2}$ s$^{-1}$ (see figure~3 of Rosati \cite{rosa98}).  In this paper,
adding {\it class 1} to {\it class 2} candidates both for patch A and
B we obtain a surface density of $\sim 22$ clusters per square degree,
out of which $\sim$ 54\% have $R = 1$ or above, which agrees well,
within the uncertainties, with the number quoted by the PDCS group,
possibly suggesting a slightly larger surface density for our
candidates, especially when taking into account the deeper magnitude
limit adopted for the search by the PDCS group. \\

We are doing multicolor (B, V, R and I) observations of the sample of
cluster candidates presented in this paper using the 3.6-m and NTT ESO
telescopes.  Redshift confirmation of three of our high redshift ($ z
\sim 0.65$) candidates has already been obtained using FORS at the
VLT, confirming the efficiency of the strategy adopted. \\

\begin{acknowledgements}
It is a pleasure to thank L. Guzzo for discussions and comments.
C. Lobo acknowledges main financial support by the CNAA fellowship
reference D.D. n.37 08/10/1997, and also the FCT PRAXIS XXI fellowship
BPD/20174/99, and the ESO/PRO/15130/1999. This research has made use
of the NASA/IPAC Extragalactic Database (NED) which is operated by the
Jet Propulsion Laboratory, California Institute of Technology, under
contract with the National Aeronautics and Space Administration.
\end{acknowledgements}

\end{document}